\documentclass[aps, pre, showpacs, twocolumn, superscriptaddress]{revtex4}
\usepackage{graphicx}
\usepackage{amsmath, amssymb}
\begin{document}
\title{Local dynamics of a randomly pinned crack front: A numerical study}
\date{\today}
\author{Knut S. \surname{Gjerden}}
\email{Knut.Skogstrand.Gjerden@gmail.com}
\affiliation{Department of Physics,
Norwegian University of Science and Technology, N--7491 Trondheim,
Norway}
\author{Arne \surname{Stormo}}
\email{Arne.Stormo@gmail.com}
\affiliation{Department of Physics,
Norwegian University of Science and Technology, N--7491 Trondheim,
Norway}
\author{Alex \surname{Hansen}}
\email{Alex.Hansen@ntnu.no}
\affiliation{Department of Physics,
Norwegian University of Science and Technology, N--7491 Trondheim,
Norway}
\begin{abstract}
We investigate numerically the dynamics of crack propagation along 
a weak plane using a model consisting of fibers connecting a soft 
and a hard clamp.  This bottom-up model has previously been shown to 
contain the competition of two crack propagation mechanisms: coalescence 
of damage with the front on small scales and pinned elastic line motion
on large scales.  We investigate the dynamical scaling properties of the 
model, both on small and large scale.  The model results compare favorable 
with experimental results. 
\end{abstract}
\maketitle
\section{Introduction}

The motion of a fracture front through a disordered medium is for
obvious reasons of great practical importance.  Nevertheless, it is 
a very complex problem which have kept materials scientists and 
mechanical engineers occupied for almost two centuries \cite{l93}.  
In an attempt to simplify the problem, Schmittbuhl et al.\ \cite{srvm95}
proposed to study the motion of a fracture front moving along a weak
plane, thereby supressing the out of plane motion of the front.  In a
seminal experimental study, Schmittbuhl and M{\aa}l{\o}y \cite{sm97}
followed this idea up by sintering two sandblasted plexiglass plates 
together and then plying them apart from one edge.  The sintering made the
otherwise due to the sandblasting, opaque plates transparent.  Where the 
plates were broken apart again, the opaqueness returned.  Hence, it was
possible to identify and follow the motion of the front since the plates
would be transparent in front of the crack front and opaque behind it.
There have been a large number of studies on this system following this
initial work. Most of these studies have been
theoretical or numerical in character and they have mostly dealt with the 
roughness of the crack front, see e.g.\ 
\cite{ref97,dsm99,rk02,shb03,sgtshm10}.
See also the recent reviews by Bonamy and Bonamy and Bouchaud \cite{b09,bb11}.

The roughness is only one aspect of the motion of the crack front in this 
system.  A first study of the dynamics of the front was published by 
M{\aa}l{\o}y and Schmittbuhl in 2001 \cite{ms01}.  The velocity
distribution was measured and found to decay slower than an exponential.
This was followed by a study by M{\aa}l{\o}y et al.\ \cite{msst06}
introducing the waiting time matrix technique making it possible to measure 
pixel by pixel the time the front sits still locally.  This revealed an
avalanche structure where portions of the front would sweep an area $S$
with a velocity $v$ before halting again.  Both the distribution of areas
and velocities were power laws.  In two later papers, Tallakstad et al.\
\cite{ttssm11} and Lenglin{\'e} et al.\ \cite{ltseatsm11}, have continued and
refined this work.  

The fluctuating line model, based on the idea that the crack front moves
as a pinned elastic line \cite{srvm95,bblp93}, dominates the theoretical
descriptions of the constrained crack problem.   It is a top-down approach
where the motion of the crack front is derived from continuum elastic
theory.  Its earliest use was to
calculate the roughness exponent controlling the scaling properties of the
crack front.  If $h(x,t)$ is the position of the crack front at time $t$
and position $x$ along its base line, then one finds the height-height 
correlation function $\langle (h(x+\Delta x)-h(x))^2\rangle$ scaling 
as $|\Delta x|^\zeta$ where $\zeta$ is the rougness exponent \cite{srvm95}.
The most presise measurement of the roughness exponent withing the 
fluctuating line model to date is $\zeta=0.388\pm0.002$ \cite{rk02}.
However, the experimental measurements of the $\zeta$ gave systematically
a much larger value, namely around 0.6 --- Schmittbuhl and M{\aa}l{o}y
\cite{sm97} found $\zeta=0.55\pm0.05$ and Delaplace et al.\ 
\cite{dsm99} $\zeta=0.63\pm0.03$.  

Santucci et al.\ \cite{sgtshm10} has reanalyzed data from a number
of earlier studies, including \cite{dsm99}, finding that the crack
front has {\it two scaling regimes:\/} one small-scale regime described
by a roughness exponent $\zeta_-=0.60\pm0.05$ and a large-scale
regime described by a roughness exponent $\zeta_+=0.35\pm0.05$.  Hence,
there is a large scale roughness regime which is consistent with the 
fluctuating line model.  However, the small scale regime is too rough.
Laurson et al.\ \cite{lsz10} has linked this regime to correlations in
the pinning strength below the Larkin lenght within the fluctuating line
model. This approach assumes that the physics behind both scaling regimes
is describable within the same top-down model.

A very different approach to explain the roughness of the crack front has been
proposed by Schmittbuhl et al.\ \cite{shb03}. It is a bottom-up approach 
based on a stress-weighted gradient percolation process.  
This idea in turn has its origin in the 
proposal by Bouchaud et al.\ \cite{bbfrr02} that the crack front does not 
advance not only due to a competition between effective elastic forces and 
pinning forces at the front, but also by {\it coalescence\/} of damage in 
front of the crack with the advancing crack itself.  The coalescence 

By refining the model proposed by Batrouni et al.\ \cite{bhs02,gsh12}, 
consisting of fibers clamped between a hard and a soft block and with a
gradient in the breaking thresholds, we have identified two scaling regimes
for the roughness of the advancing crack front \cite{gsh13}.  On large scales
we recover the roughness seen in the reanalysis of experimental data by
Santucci et al.\ \cite{sgtshm10}, $\zeta=0.39\pm0.04$, consistent with the
fluctuating line model, whereas on small scales with identify a gradient
percolation process, leading to $\zeta=2/3$ \cite{hbrs07}, which is also
consistent with Santucci et al.\ reanalysis. 

Bonamy et al.\ \cite{bsp08} and Laurson et al.\ \cite{lsz10} have considered
the crackling dynamics \cite{sdm01} of the fluctuating line model finding
behavior consistent with the experimental measurements
\cite{ms01,msst06,ttssm11,ltseatsm11}.  

We study in this paper the crackling dynamics in the fiber bundle model
of Gjerden et al.\ \cite{gsh12,gsh13}.  We find quantitative consistency 
with the experimental measurements.  Hence, we demonstrate that this simple
model contains the observed experimental features of constrained crack growth.
It contains the essential physics of the problem.  Since we have control
over the crossover length scale between the crack advancing due to crack
coalescence and due to pinning of the crack front as an elastic line, we
analyse our results in light of this.

In Section \ref{sec1}, we describe the model in detail. Section \ref{sec2}
continues the analysis from Ref.\ \cite{gsh13} of the roughness of the 
crack front.  In \cite{gsh13}, we measured the roughness locally using 
the average wavelet coefficient method \cite{mrs97,shn98}.  Here we base
our analysis on the variance of the front, again finding two regimes: a 
small-scale regime consistent with ordinary gradient percolation and a 
large-scale regime consistent with the fluctuating line model.  In
Section \ref{sec3} we examine the velocity distribution during the motion
of the crack front.  Our data are consistent with the findings of
M{\aa}l{\o}y et al.\ \cite{msst06} and Tallakstad et al.\ \cite{ttssm11}.
Section \ref{sec4} is devoted to an analysis of the geometry of the 
avalanches that occur during the motion of the crack front.  The last 
section before the conclusion, Section \ref{sec5} contains a tying up of
loose ends through a discussion of the results.  

The main aim of this paper has been to demonstrate that the simple model
we use is capable of reproducing the experimental data.  Hence, this simple
fiber bundle model seems to contain the essential physics of the
constrained crack problem.  

\section{Model}
\label{sec1}

The model we base our calculations on is a refinement of the fiber
bundle model \cite{phc10} used by Schmittbuhl et al.\ \cite{shb03}
and introduced by Batrouni et al.\ the year before \cite{bhs02}.
$L\times L$ elastic fibers are placed in a square lattice between two
clamps.  One of the clamps is infinely stiff whereas the other has a finite
Young modulus $E$ and a Poisson ratio $\nu$.  All fibers are equally long
and have the same elastic constant $k$.  We measure the position of the stiff
clamp with respect its position when all fibers carry zero force, $D$.  The
force carried by the fiber at position $(i,j)$, where $i$ and $j$ are
coordinates in a cartesian coordinate system oriented along the
edges of the system, is then
\begin{equation}
\label{base}
f_{(i,j)}=-k(u_{(i,j)}-D)\;,
\end{equation}
where $u_{(i,j)}$ is the elongation of the fiber at $(i,j)$ and $k$ is the 
spring constant of the fibers.  It is assumed to be the same for all fibers.

The fibers redistribute the forces
they carry through the response of the clamp with finite elasticity.
The redistribution of forces is accomplished by using the Green function
connecting the force $f_{(m,n)}$ acting on the clamp from fiber
$(m,n)$ with the
deformation $u_{(i,j)}$ at fiber $(i,j)$, \cite{j85}
\begin{subequations}
\label{greena}
\begin{align}
u_{(i,j)}&= \sum_{(m,n)}G_{(i,j),(m,n)}f_{(m,n)},\\
G_{(i,j),(m,n)}& = \nonumber \\
\frac{1-\nu^2}{\pi E a^2}&\int^{a/2}_{-a/2} dx\ \int^{a/2}_{-a/2} dy\
\frac{1}{|\vec{r}_{(i,j)}-\vec{r}_{(m+x,n+y)}|}. \label{greenb}
\end{align}
\end{subequations}
where $a$ is the distance between neighboring fibers, $\nu$ the Poisson 
ratio and $E$ the elastic constant. The integration in Eq.\ \ref{greenb}
is performed over the voronoi cells around each fiber.
$\vec{r}_{(i,j)}(0,0)-\vec{r}_{(m,n)}(0,0)$ is the distance 
between fibers $(i,j)$ and $(m,n)$.

The equation set, Eqs.\ (\ref{greena}) and (\ref{greenb}), is solved using a 
Fourier accelerated conjugate gradient method \cite{bhn86,bh88}.

The Green function, Eq.\ (\ref{greenb}), is proportional to
$(Ea)^{-1}$.  The elastic constant of the fibers, $k$, must be
proportional to $a^2$.  The linear size of the system is $aL$.  Hence,
by changing the linear size of the system without changing the discretization
$a$, we change $L \to \lambda L$ but leave $(Ea)$ and $k$ unchanged.
If we on the other hand change the discretization without changing the
linear size of the system, we simultaneously set $L\to\lambda L$,
$(Ea)\to\lambda (Ea)$ and $k\to k/\lambda^2$. We define the scaled
Young modulus $e=(Ea)/L$.  Hence, changing $e$ without changing $k$ is
equivalent to changing $L$ --- and hence the linear size of the system ---
while keeping the elastic properties of the system constant \cite{sgh12}.

The fibers are broken by using the quasistatic approach \cite{h05}.  That is,
we assign to each fiber $(i,j)$ a threshold value $t_{(i,j)}$.
They are then broken one by one by each time identifying
$\max_{(i,j)}(f_{(i,j)}/t_{(i,j)})$ for $D=1$. This ratio
is then used to read off the value $D$ at which the next fiber breaks.

In the constrained crack growth experiments of Schmittbuhl and M{\aa}l{\o}y
\cite{sm97}, the two sintered plexiglass plates were plied apart from one
edge.  In the numerical modeling of Schmittbuhl et al.\ \cite{shb03},
an asymmetric loading was accomplished by introducing a linear
gradient in $D$. Rather than implementing an asymmetric loading, we
use a gradient in the threshold distribution,
$t_{(i,j)}=g j + r_{(i,j)}$, where $g$ is the gradient and $r_{(i,j)}$
is a random number drawn from a flat distribution on the unit interval.
In the limit of large Young modulus $E$, i.e., both clams are stiff,
the Green function vanishes and the system becomes equivalent to
the gradient percolation problem \cite{srg85,gr05}.

In order to follow the crack front as the breakdown process develops, we
implement the ``conveyor belt" technique \cite{drp01,gsh12} where a new upper
row of intact fibers is added and a lower row of broken fibers removed
from the system at regular intervals.  This makes it possible to follow
the advancing crack front indefinitely.  The implementation has been
described in detail in \cite{gsh12}.
\section{Roughness and dynamical scaling}
\label{sec2}

We demonstrate in this Section the existence of two scaling regimes for the 
roughness of the crack front by providing evidence beyond that which was
presented in Gjerden et al.\ \cite{gsh13}.

We identify the crack front by first eliminating all islands of
surviving fibers behind it and all islands of failed fibers in front
of it.  We measure ``time" $n$ in terms of the number of failed fibers.
After an initial period, the system settles into a steady state. We then
record the position of the crack front $j=j(i, n = 0)$ after having set
$n=0$.  We then define the position at later times $n>0$ relative to this
initial position,
\begin{equation}
\label{defh}
h_i(n)=j(i,n)-j(i,n=0)\;.
\end{equation}
This is the same definition as was used by Schmittbuhl and M{\aa}l{\o}y
\cite{sm97}. The front as it has now been defined will contain overhangs.
That is, there may be multiple values of $h_i(n)$ for the same $i$ and $n$
values.  We only keep the largest $h_i(n)$, i.e.\ we implement the 
Solid-on-Solid (SOS) front.

We define the average position of the front as 
\begin{equation}
\label{aveh}
\langle h(n)\rangle=\frac{1}{L}\ \sum_{i=1}^L h_i(n)\;,
\end{equation} 
and the front width as
\begin{equation}
w(n)^2=\frac{1}{L}\ \sum_{i=1}^L(h_i(n)-\langle h(n)\rangle)^2\;.  
\end{equation}

\begin{figure}[ht]
\includegraphics[width=3.5in]{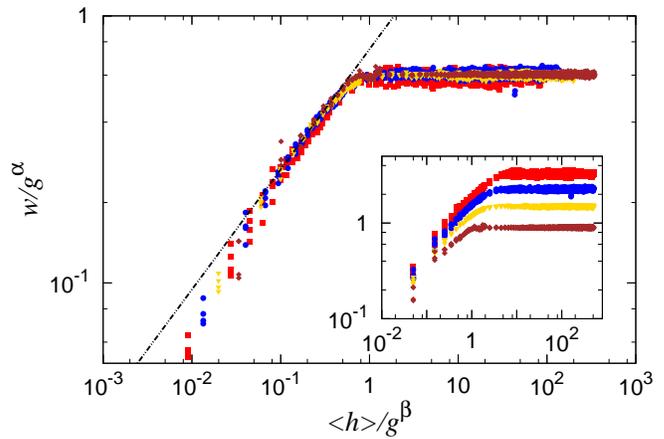}
\caption{(color online). One-parameter scaling of the standard deviation 
of the width of the front as a function of the average position of the 
front as the front moves forward from some initial (rough) configuration. 
The main figure shows data collapse for $\alpha=\beta=4/7$, and the 
inset shows the same data with no scaling, i.e. $\alpha=\beta=0$. 
The straight line has slope $288/637\approx 0.45$ --- see text.
The data are extracted from simulations with, from top-down in the 
inset, $g\in(0.05,0.1,0.2,0.5)$. Each group in the inset is comprised 
of data from systems of sizes 64-256 with constant $E/L$.}
\label{fig1}
\end{figure}

We start by examining the front in the range of values of the scaled
elastic constant $e$ for which the front has the character of a 
gradient percolation hull, $e=3.1$ \cite{gsh13}.  
Fig.\ \ref{fig1} shows one-parameter scaling of the front width $w$ 
versus the average height of the 
front $\langle h\rangle$ with the gradient $g$ andas the scaling parameter. 
There is data collapse for a range of system sizes and gradients. 
Based on measuring the fractal dimension of the front and using 
the average wavelet coefficient method to determine the roughness exponent,
we concluded in \cite{gsh13} that the front roughness is in the ordinary
gradient percolation universality class.  This implies that the two
scaling exponents in Fig.\ \ref{fig1} are $\alpha=\beta=\nu(1+\nu)=4/7$,
where $\nu=4/3$, the percolation value \cite{hbrs07}.

We now assume that the front width obeys Family-Vicsek scaling \cite{fv85},
i.e., 
\begin{equation}
\label{fvscaling}
w(L,\langle h\rangle) = L^\zeta\ 
f\left(\frac{\langle h\rangle}{L^\kappa}\right)\;,
\end{equation}
where 
\begin{equation}
\label{fvfunc}
f(q) = \left\{
\begin{array}{l l}
q^{\zeta/\kappa} & \quad \text{if $q\to 0$\;, }\\
constant & \quad \text{if $q\to\infty$\;.}\\
\end{array} \right.
\end{equation}
For $\langle h\rangle \ll L^\kappa$, and assuming that we are dealing with
ordinary percolation, we may relate $\langle h\rangle$ to the number of
fibers below and including 
the front, $A$, by the fractal dimension $D_A=91/48$ \cite{sa92},
$\langle h\rangle^{D_A} \sim A$.  Invoking Mandelbrot's area-perimeter
relation, we may also relate the length of the front, $l$ to the number of
fibers below and including the front, $l\sim A^{D_h/2}$, where $D_h=7/4$
\cite{gr05, m77}.  Eliminating $A$ between these two expressions leads to
$l\sim \langle h\rangle^{D_H D_A/2}$.  We relate the length of the front
$l$ to its width $w$ by the relation $l \sim w^{D_H} (L/w)$ \cite{gsh13}. 
Eliminating $l$ between these two relations gives
\begin{equation}
\label{wvshsmall}
w \sim \langle h\rangle^{D_A D_H /(2D_H-2)}= \langle h\rangle^{288/637}\;,
\end{equation}
where $288/637\approx 0.45$.  The straight line in Fig.\ \ref{fig1} has
this slope.  We may now calculate the dynamical exponent $\kappa$ in
Eq.\ (\ref{fvscaling}).  We find $\kappa=637/432\approx 1.47$.

\begin{figure}[ht]
\includegraphics[width=3.5in]{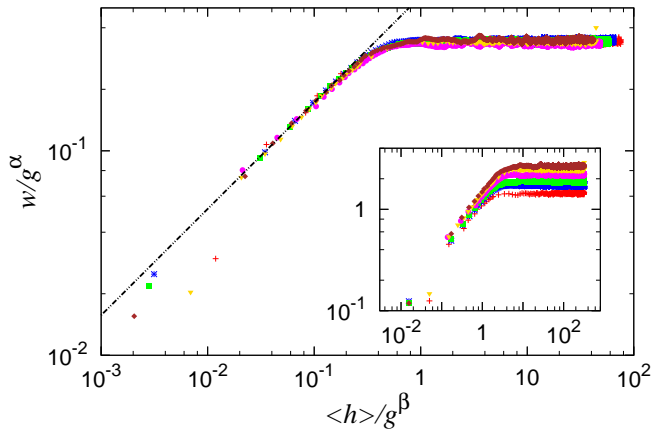}
\caption{(color online). One-parameter scaling of $w$ as a function of 
$\langle h\rangle$ with $g$ as the scaling parameter. The main 
figure shows data collapse for $\alpha=\beta=0.4$. The inset shows 
unscaled data, i.e. $\alpha=\beta=0$.  The slope of the straight line
is $0.52$.  The parameter values are, 
from top-down in the inset, $g\in(0.007,0.008,0.01,0.015,0.02,0.03)$. 
The system size is $L=64$.}
\label{fig2}
\end{figure}

Fig.\ \ref{fig2} shows the corresponding one-parameter scaling as in 
Fig.\ \ref{fig1}, but now for systems with a much lower 
$e=7.8\times10^{-4}$, placing us in the realm of the fluctuating
line model \cite{gsh13}.  The two scaling exponents that produce
data collapse are $\alpha=\beta=0.4$.  By a least squares fit, we determine
$w\sim \langle h\rangle^{0.52\pm0.05}$.  This value is very close to the 
one seen for large $e$ values, and consistent with the measurement of 
Tallakstad et al.\ \cite{ttssm11}, who found $0.55$ and Schmittbuhl et al.\
\cite{sdm01-1} who found $0.52$, both groups using sintered PMMA plates.

Assuming Family-Vicsek 
scaling, we determine the dynamical exponent to be 
$\kappa=0.75\pm0.07$.  This is consistent with the value reported by Duemmer
and Krauth for the fluctuating line model, $\kappa=0.70\pm 0.5$.

The dynamical exponent reported by M{\aa}l{o}y and Schmittbuhl \cite{ms01}
for the PMMA system
was $\kappa=1.2$.  We find $\kappa=1.47$ on small scales and $\kappa=0.75$
on large scales.  

\begin{figure}[ht]
\includegraphics[width=3.5in]{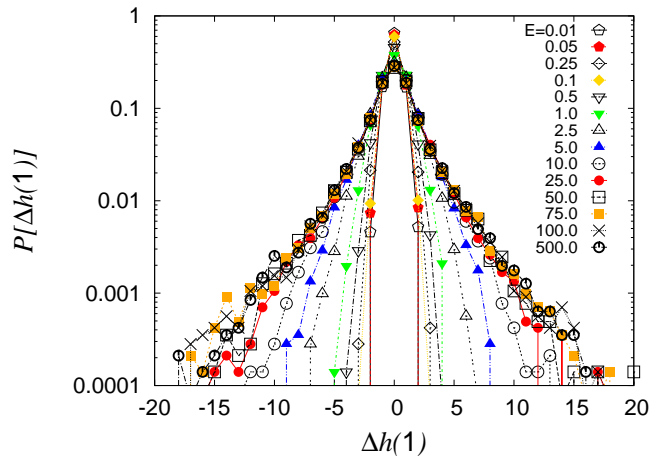}
\caption{(color online). Histogram of height fluctuations with respect to
$\Delta h(1)=|h_{i+1}(n)-h_i(n)|$ for different elastic constants. The
distribution follows a gaussian when the system is soft and broadens when
the system stiffens.}
\label{fig2-1}
\end{figure}

It was observed by Santucci et al.\ \cite{sgtshm10} that the large-scale
roughness regime shows a gaussian distribution of the height fluctuations,
whereas in the small-scale it broadens to a non-gaussian distribution.  We 
investigate this in our model.  We define the height fluctuations as
$\Delta h(1)=|h_{i+1}(n)-h_i(n)|$.  By changing the elastic constant,
$E$ in our model while keeping the step size constant, the result is equivalent
to changing the step size in the experiment.  We show in Fig.\ \ref{fig2-1}
the distribution of fluctuations for different $E$.  We see that for soft
systems, the distribution is gaussian.  As the system stiffens, the 
distribution broadens as in the experiment.

\section{Local velocity distribution}
\label{sec3}

\begin{figure}[ht]
\includegraphics[width=3.5in]{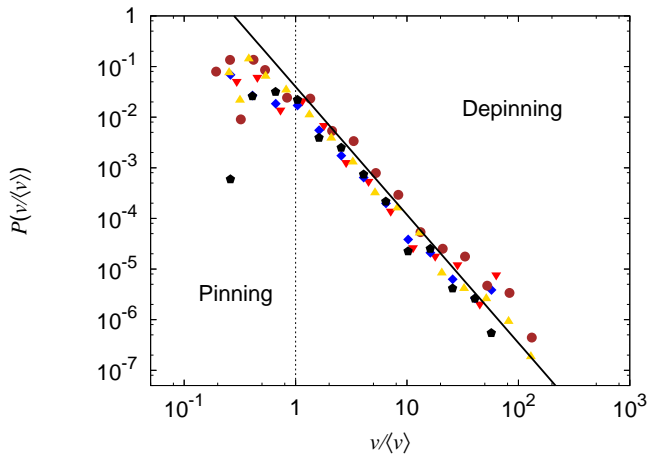}
\caption{(color online). Distribution of local velocities scaled by the 
global average velocity. Results are based on simulations of sizes 
$L=64,128$ with $e=1.56\times10^{-4}, 7.81\times10^{-4}$, 
respectively, and loading gradient in the range $g\in[0.01-0.05]$. 
A fit to the data for $v>\langle v\rangle$ yields a power law 
behavior with an exponent of $-2.53$.}
\label{fig3}
\end{figure}

We examine the distribution of local velocities using the 
waiting time matrix (WTM) method \cite{msst06}. The WTM method 
consists of letting the fracture move across a matrix with all elements
equal to zero in discrete time steps. Every time step, the matrix elements 
that contain the front are incremented by one.
When the front has passed, the matrix contains the time the front spent in
each position.  Hence, matrix elements contaning low values correspond 
to rapid front movements and large values correspond to slow, pinned movement. 
From the WTM, one can then calculate the spatiotemporal map of velocities 
by mapping the coordinates of the front at each step in the simulation 
onto the local velocities $V(x,y)$ calculated from the inverse of 
the waiting time, thus obtaining $v(x,t)$. From this map we then 
calculate the global average velocity $\langle v \rangle$ and the 
distribution of local velocities $P(v)$.  It has been found 
experimentally that this distribution follows 
\begin{equation}
\label{veldist}
P(v/\langle v\rangle) 
\sim (v/\langle v\rangle)^{-\eta}, 
\quad \textrm{for} \quad v/\langle v\rangle > 1,
\end{equation}
with the exponent $\eta=2.55\pm0.15$ \cite{msst06,ttssm11}. The 
results from our simulations are shown in Fig.\ \ref{fig3} for
scaled elastic constant $e=1.6\times 10^{-4}$ and $7.8\times 10^{-4}$.
The slope in the figure is obtained through a linear fit and has the 
value $2.53$.  Based on this figure, we follow Tallakstad et al.\ 
\cite{ttssm11} and  define a \textit{pinning} regime where 
$v<\langle v\rangle$ and a \textit{depinning} regime where 
$v>\langle v\rangle$.  

\subsection{Space and time correlations}
\label{sub3.1}

\begin{figure}[ht]
\includegraphics[width=3.5in]{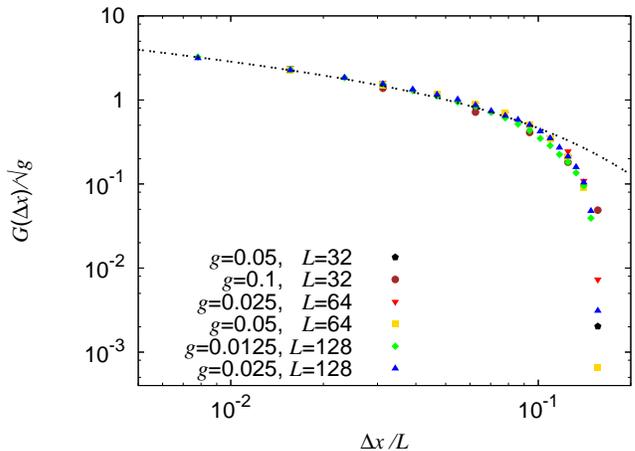}
\caption{(color online). Spatial correlation function $G(\Delta x)$. 
Data collapse is obtained when $\Delta x$ is scaled with the system 
size and $G(\Delta x)$ is scaled with $\sqrt{g}$. The line is a fit 
to a power law with exponential cutoff 
$G(\Delta x)=A (\Delta x/L)^{-\tau_x}\textrm{exp}(-\Delta x/0.1L)$ 
with $\tau_x=0.4$. The parameters used are as in Fig.\ \ref{fig3}.
}
\label{fig4}
\end{figure}

From the local velocities, we calculate the normalized correlation 
functions in space, $G(\Delta x)$, and time, $G(\Delta t)$ 
\begin{subequations}
\label{green}
\begin{align}
G(\Delta x) &= \left\langle 
\frac{\langle[v(x+\Delta x,t)-\langle v \rangle_x]
[v(x,t)-\langle v \rangle_x]\rangle}{\sigma_x^2} \right\rangle_t\;,\\
G(\Delta t) &= 
\left\langle \frac{\langle[v(x, t+\Delta t)-\langle v 
\rangle_t][v(x,t)-\langle v \rangle_t]\rangle}{\sigma_t^2} \right\rangle_x\;.
\end{align}
\end{subequations}
$\sigma_{x/t}$ and $\langle\phantom{v}\rangle_{x/t}$ denote standard 
deviation and average over $x$ or $t$. The results for $G(\Delta x)$ 
is presented in figure \ref{fig4}. The results are in qualitative 
agreement with what has been reported experimentally \cite{ttssm11}. 
Due to the small system sizes available numerically, the cutoff we observe 
is extremely sharp. Still, the data appears to be consistent with a power law
distribution with an exponential cutoff 
$G(\Delta x)=A (\Delta x/L)^{-\tau_x}\textrm{exp}(-\Delta x/0.1L)$ 
with $\tau_x=0.4$. This is in the same range as the values
reported by Tallakstad et al.\ \cite{ttssm11}, where $\tau_x=0.53\pm0.12$.

\begin{figure}[ht]
\includegraphics[width=3.5in]{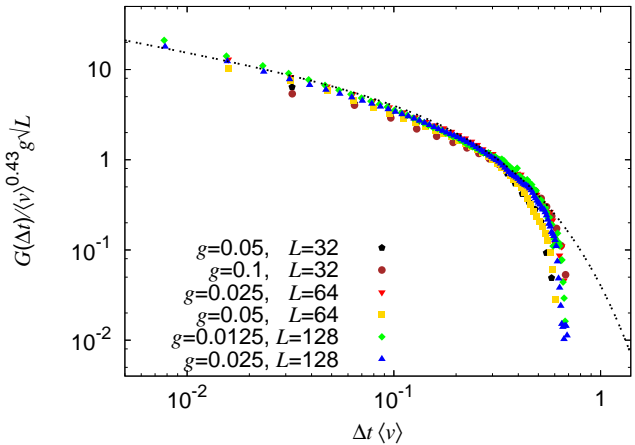}
\caption{(color online). Temporal correlation function. Best data 
collapse is obtained when scaling $\Delta t$ by 
$1/\langle v \rangle$ and $G(\Delta t)$ by 
$\langle v \rangle^{0.43}g\sqrt{L}$. The line is a fit to a 
power law with exponential cutoff 
$G(\Delta t)=B (\Delta t\langle v \rangle)^{-\tau_t}
\textrm{exp}(-\Delta t\langle v \rangle/0.25)$ with $\tau_t=0.43$.
The parameters used are as in Fig.\ \ref{fig3}.
}
\label{fig5}
\end{figure}

In figure \ref{fig5}, we show the results of the analysis of the 
temporal velocity distribution. Again, we attempt to fit the data to a power 
law with exponential cutoff, $G(\Delta t)=B 
(\Delta t\langle v \rangle)^{-\tau_t}\textrm{exp}
(-\Delta t\langle v \rangle/0.25)$, with exponent $\tau_t=0.43$. 
The scaling resemble closely that observed by Tallakstad et al.\
\cite{ttssm11}. 

We note that data collapse is obtained in Fig.\ \ref{fig4}
when the correlation function is scaled by $g^{1/2}$ and $\Delta x$ by $L$, 
and in Fig.\ \ref{fig5}, the correlation function is scaled by $gL^{1/2}$.

\section{Cluster analysis}
\label{sec4}

In the following we analyze the geometrical structure of the areas
swept by the crack front during avalanches.

\begin{figure}[ht] 
Depinning \qquad\qquad\qquad\qquad Pinning \\
C=2 \qquad\qquad\qquad\qquad\qquad C=1.5 \\
\includegraphics[width=4.2cm]{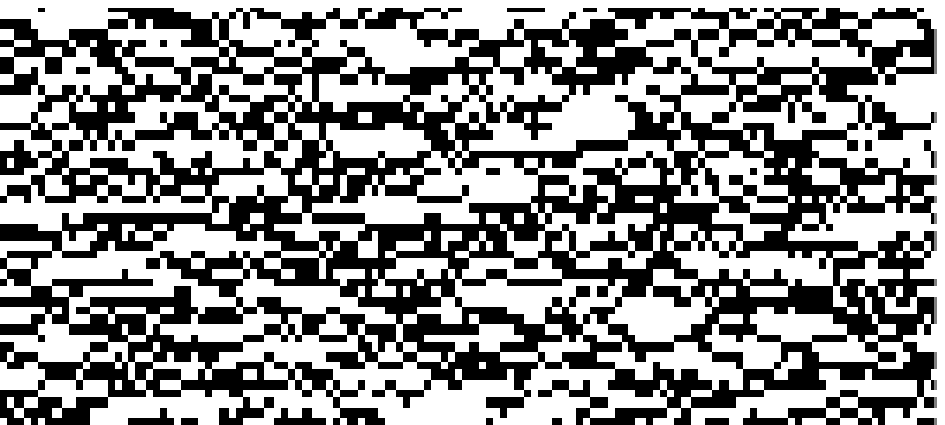} 
\includegraphics[width=4.2cm]{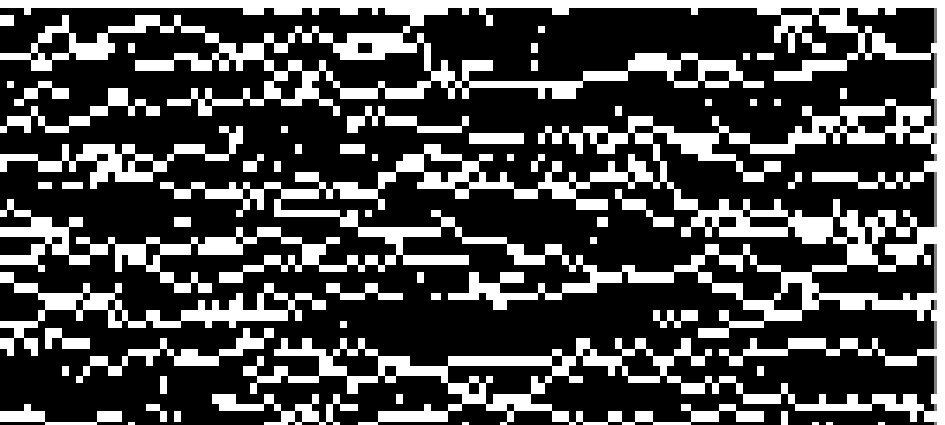} \\
C=4 \qquad\qquad\qquad\qquad\qquad C=3 \\
\includegraphics[width=4.2cm]{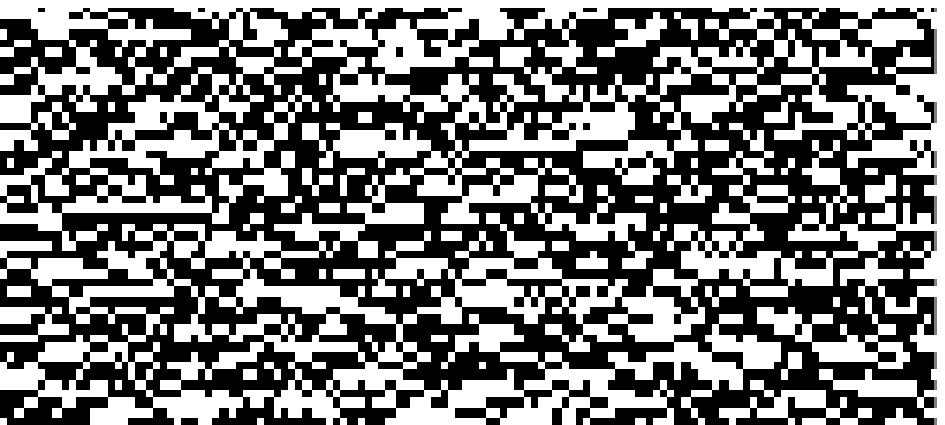} 
\includegraphics[width=4.2cm]{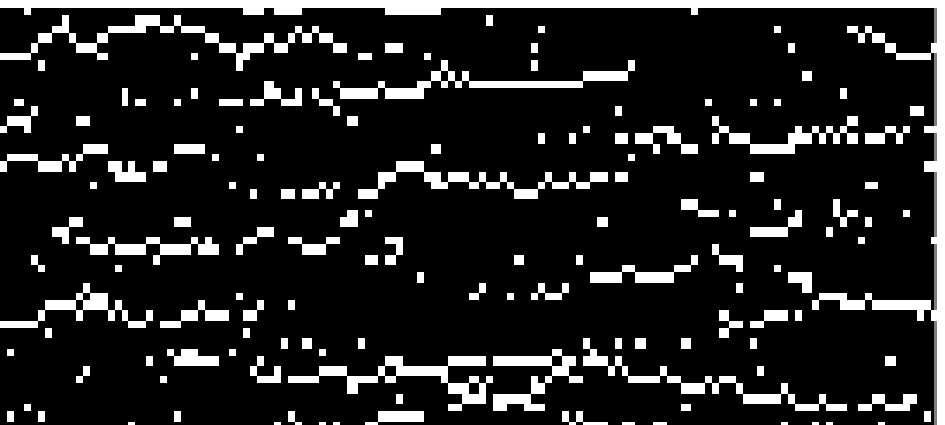}  \\
C=10 \qquad\qquad\qquad\qquad\qquad C=5 \\
\includegraphics[width=4.2cm]{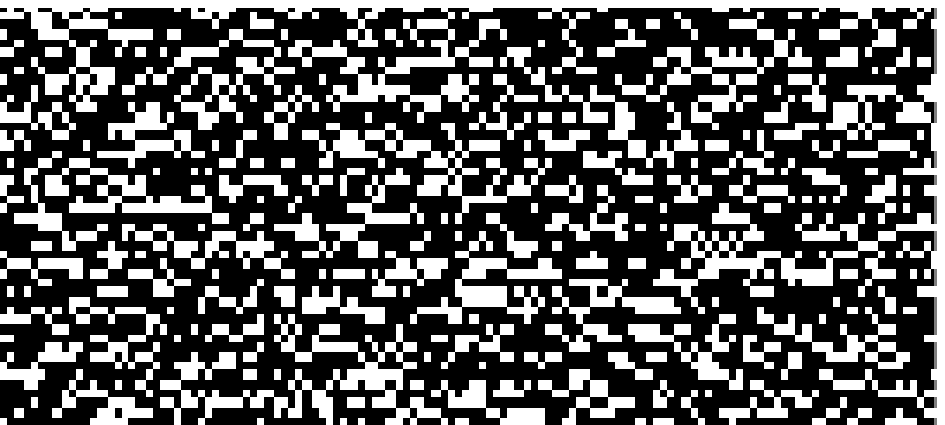} 
\includegraphics[width=4.2cm]{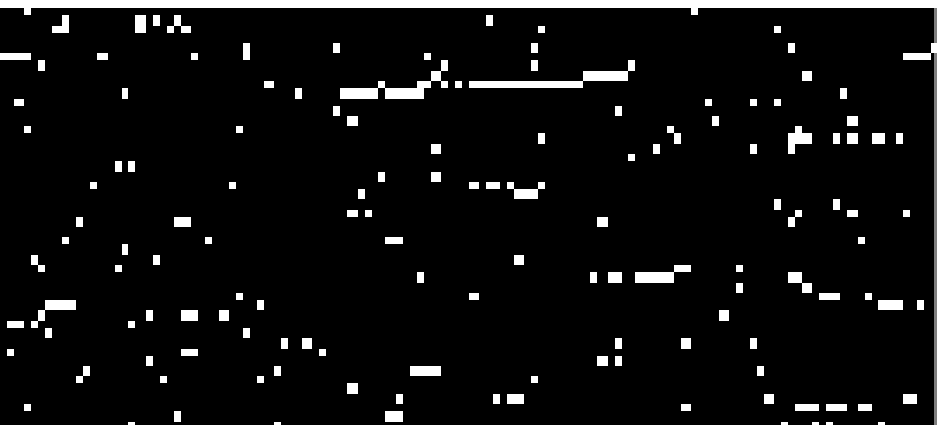} 
\caption{Part of the thresholded velocity matrix showing 
depinning clusters on the left and pinning clusters on the 
right. The images are from a system of size $L=128$, $g=0.0125$ and 
$e=7.8\times10^{-4}$. White areas are areas where the local 
velocities are either above C$\langle v \rangle$ (depinning) 
or below $\frac{1}{\textrm{C}}\langle v \rangle$ (pinning). 
}
\label{fig6}
\end{figure}

Following Tallakstad et al.\ \cite{ttssm11}, we construct thresholded
velocity maps $V_C$ in such a way that 
\begin{equation}
V_C = \begin{cases} 1, & \mbox{for } v 
\geq C\langle v \rangle\\ 0, & \mbox{for } 
v < C\langle v \rangle \end{cases}\;,
\end{equation}
in the depinning regime and
\begin{equation}
V_C = \begin{cases} 1, & \mbox{for } 
v \leq \langle v \rangle/C\\ 0, & \mbox{for } 
v > \langle v \rangle/C \end{cases}\;,
\end{equation}
in the pinning regime. $C$ is a constant in the range of $2-12$ for 
depinning and $1.5-6$ for pinning for our simulations. For larger 
system sizes, larger $C$ are possible. Examples of these matrices 
are shown in figure \ref{fig6}, where the structural difference 
between depinning and pinning clusters can be seen. 
In the spirit of de Saint-Exup{\'e}ry we may characterize the 
pinning clusters as snakelike, whereas the depinning clusters are
elephant-in-snakelike \cite{e43}.
The following analysis uses only $L=64$
and $L=128$, and are based on 250 and 100 samples, respectively.
$g=0.025$ for the smaller systems, and $g=0.0125$ for the larger systems. 
For $L=64$, $e=1.6\times10^{-4}$, and for $L=128$, 
$e=7.8\times10^{-4}$.
The systems will for these $e$-values behave virtually 
identical \cite{gsh13}.

\subsection{Size distribution of clusters}
\label{sub4.1}

We denote the number of bonds in a cluster the size or area of a 
cluster $S$.  Both experiments \cite{msst06,ttssm11} 
and numerical simulations using the fluctuating line model 
\cite{bsp08} on in-plane fracture, 
show a power law distribution $P(S)\sim S^{-\gamma}$ for $S$
with exponent close to $\gamma=1.56$ \cite{ttssm11} or 
$\gamma=1.65$ \cite{bsp08}. 

\begin{figure}[ht]
\includegraphics[width=3.5in]{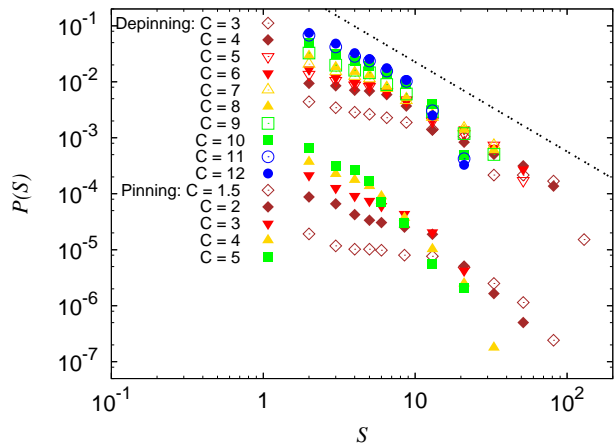}
\caption{(color online). Probability distribution function of the 
size of clusters, $S$, for a range of $C$ values for both pinning (lower
cluster of points) and depinning clusters (upper cluster of points). 
The data for pinning clusters are shifted down by 
0.005 to visually separate the two regimes. The slope of the 
line is $\gamma=1.6$.  $L=64$, $e=1.6\times10^{-4}$.
}
\label{fig7}
\end{figure}

\begin{figure}[ht]
\includegraphics[width=3.5in]{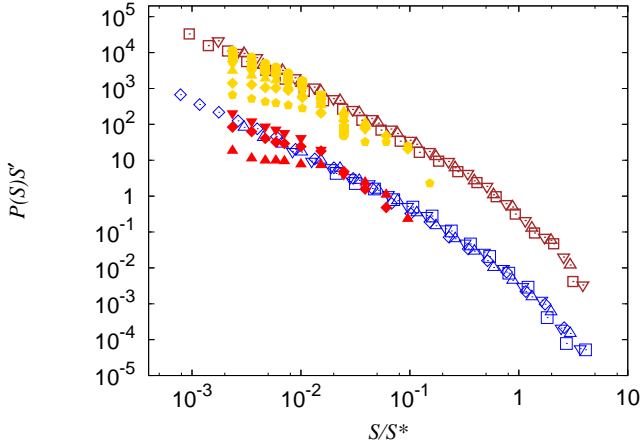}
\caption{(color online). Size distribution of clusters. Depinning 
clusters on top and pinning clusters below. The longest data 
series (unfilled markers) are experimental data obtained with 
permission from the authors \cite{ttssm11}.  The shorter data 
series (filled markers) are numerical data from Fig.\ \ref{fig7}. 
The experimental data are unchanged from \cite{ttssm11}, 
so $S'=S^{*\gamma}$ and $S^*$ are as given in 
\cite{ttssm11}. Due to lack of direct physical 
parameter comparison, the numerical data are scaled 
to match the experimental data to verify similar behavior.
}
\label{fig8}
\end{figure}

We show in Fig.\ \ref{fig7} the probability distribution of $S$
both for pinning and depinning for scaled elastic constant $e=1.6\times10^{-4}$.   
Our data are consistent with 
previous measurements, but due to the limited size of the 
numerical simulations currently available, the exponent cannot 
be accurately determined. One prominent feature of the clusters 
is that due to the small system size, the presence of large clusters 
for low values of $C$ suppresses the number of smaller clusters, 
and for high values of $C$, the number of large clusters is 
reduced by the filtering process in $C$. This is a pure finite 
size effect, which we assume will disappear for larger system sizes. 
The experiments, which would correspond to a very large 
simulation size, shows no such dependence on $C$. 
A direct comparison between the experimental and our data 
is given in Figure \ref{fig8}.

\subsection{Aspect ratio of clusters}
\label{sub4.2}

Next, we extract the linear extensions as length and height from the 
cluster sizes. This is done by enclosing each cluster in a bounding 
box of size $l_x$ by $l_y$. Although the appearance differs greatly 
between pinning and depinning clusters, the method for determining 
$l_x$ and $l_y$ are the same. As seen qualitatively in Fig.\ \ref{fig6}, 
pinning clusters are much 
longer than they are wide, and in a worst-case scenario they could 
approximate the diagonal in the bounding box, which would lead to a 
gross over-estimation of $l_y$. However, such behavior was not seen 
for $L=64,128$ and we judge the approximation of $l_y$ to be equally 
sound in both regimes.

\begin{figure}[ht]
\includegraphics[width=3.5in]{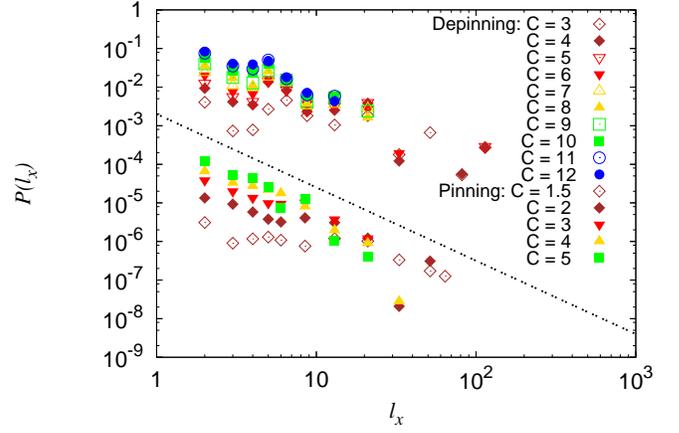}
\includegraphics[width=3.5in]{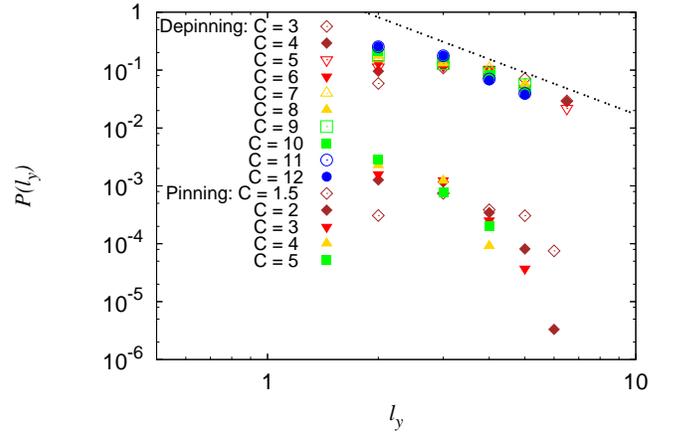}	
\caption{(color online). Probability distribution function of the 
linear extension of the clusters both parallel and perpendicular 
to the fracture front. The data for pinning clusters are shifted down 
for visual clarity by 0.001 for $l_x$ and 0.005 for $l_y$. 
The slope 
of the line in the top figure is $\beta_x=1.9$ and in the bottom 
figure $\beta_y=2.4$. 
}
\label{fig9}
\end{figure}

Figure \ref{fig9} shows the size distributions of $l_x$ and $l_y$ for
scaled elastic constant $e=1.6\times10^{-4}$. As in 
Fig.\ \ref{fig7}, there is a pronounced dependence on $C$. The data
indicate power law distributions 
$P(l_x)\sim l_x^{-\beta_x}$ and 
$P(l_y)\sim l_y^{-\beta_y}$ \cite{ttssm11}. 
The data are plotted with lines showing values of the exponents equal
to the experimentally obtained $\beta_x=1.93$ and $\beta_y=2.36$ 
\cite{ttssm11}. Both pinning and depinning data show behaviour 
consistent with the experimental data.

\begin{figure}[ht]
\includegraphics[width=3.5in]{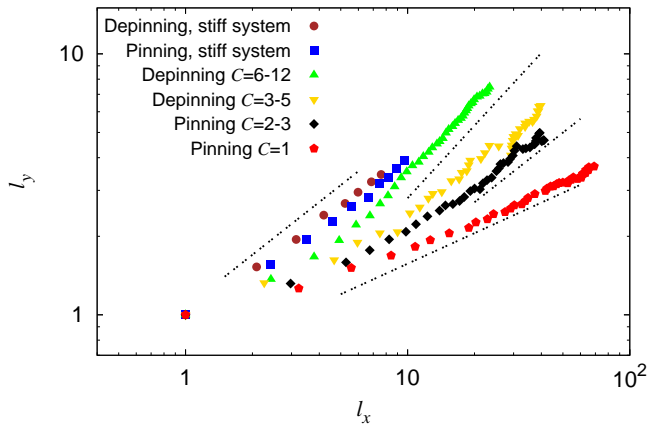}
\caption{(color online). Local scaling of the clusters. The lines 
are guides to the eye, but the slopes are based on linear regression 
and the values are in descending order 0.91, 0.67, 0.67, and 0.39. 
The data labeled stiff system are from simulations using a Young's modulus 
four orders of magnitude greater than what is used in the rest of the 
simulations. $L=128$, $g=0.0125$, and $e=7.8\times10^{-4}$ 
and $e=7.8$. Further details in the text. 
}
\label{fig10}
\end{figure}

Turning to examine the relation between $l_y$ and $l_x$, we plot 
these against each other in Fig.\ \ref{fig10}. This data is more 
well-behaved than the probability distributions, so scaling 
relations are more pronounced. The data are consistent with a relation
of type
\begin{equation}
\label{bonamyrel}
l_y\sim l_x^H\;,
\end{equation}
as studied both by Tallakstad et al.\ \cite{ttssm11} and Bonamy et al.\
\cite{bsp08} in respectively the experimental system and with the fluctuating
line model. 

We see a clear dependency of $H$ on parameters such as
$e$ and $C$. 
Considering the depinning case first for $e=7.8\times10^{-4}$, we divide the data
between $C^{\textrm{high}}\in[6-12]$, a 
group containing data from only the highest local velocities, 
and $C^{\textrm{low}}\in[3-5]$, containing lower 
local velocities.  There is a clear visual difference in behavior 
between the two groups:  The group with the higher $C$ values, have the
higher slope. We measure the corresponding exponents to be
$H_{\textrm{d}}^{\textrm{high}}=0.9$ and 
$H_{\textrm{d}}^{\textrm{low}}=0.65$. 

In the pinning regime we similarly divide between low and high 
local velocity effects, except that the range of available $C$ is 
much smaller, so we find $C^{\textrm{high}}=2,3$ and 
$C^{\textrm{low}}=1$. This yields the two exponents 
$H_{\textrm{p}}^{\textrm{high}}=0.65$ and 
$H_{\textrm{p}}^{\textrm{low}}=0.4$. 

It has been suggested that $H$ is another measure of $\zeta$, the 
roughness exponent \cite{msst06,bsp08,lsz10}.  In order to test this, we
have added data both for pinning and depinning of
a stiff system ($e=7.8$) in Fig.\ \ref{fig10}.  If the conjecture is
right in this case, we would expect a $H$ exponent equal to $2/3$.  We have
added such slopes, and indeed this seems fulfilled.

We may then speculate whether the exponents that we measure for the softer
system ($e=7.8\times10^{-4}$), indeed are roughness exponents --- at least for 
some groups of $C$ values.  

\begin{figure}[ht]
\includegraphics[width=3.5in]{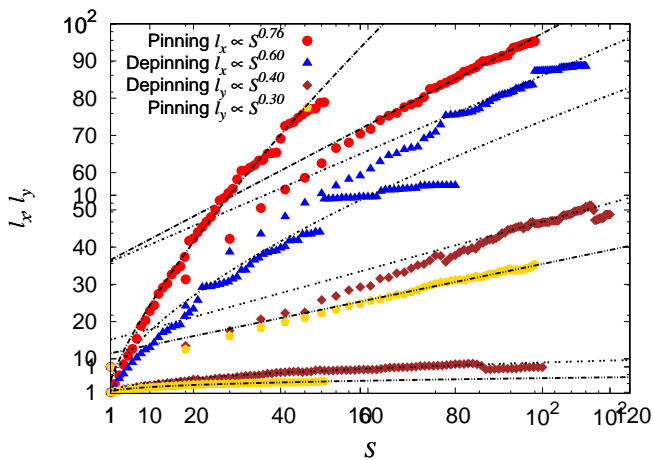}
\caption{(color online). Linear size of both pinning and depinning 
clusters as a function of cluster size. The simulation parameters 
are the same as for Fig.\ \ref{fig10}.
}
\label{fig11}
\end{figure}

Another way to examine these scaling relations is to check the 
relation between $l_x, l_y$ and $S$. We expect to see a power 
law relationship of type 
\begin{equation}
\label{checklxly}
l_i\sim S^{\alpha_i}\;.
\end{equation}
The data are plotted in Fig.\ \ref{fig11}, in which data for the 
highest local velocities are not included, $C=C^{\textrm{low}}$ 
for both pinning and depinning. In the depinning regime, we 
find the exponents $\alpha_{x}^{\textrm{d}}=0.6$ 
and $\alpha_{y}^{\textrm{d}}=0.4$. Eliminating $S$, 
we get a measure of $H$ through 
\begin{equation}
\label{check2}
l_y\sim l_x^{\alpha_y/\alpha_x}=l_x^H\;,
\end{equation}
resulting in $H_{\textrm{d}}=0.65$. In the pinning regime 
we find $\alpha_{x}^{\textrm{p}}=0.75$ and 
$\alpha_{y}^{\textrm{p}}=0.3$, yielding 
$H_{\textrm{p}}=0.4$. 

These results are to be compared to the experimental measurements
\cite{ttssm11} giving $\alpha_x=0.62\pm0.04$ in pinning and
the depinning regime and $\alpha_y=0.41\pm0.06$ or $0.34\pm0.05$
in the depinning regime depending on how the bounding box was
defined.  This leads to the value $H=\alpha_y/\alpha_x=0.66$ or 0.55
depending on the bounding box used.  Bonamy et al.\ \cite{bsp08}
found $H=0.65\pm0.05$.

\section{Discussion}
\label{sec5}

In our model, $e$ and $g$ are the most important input parameters. 
$g$ effectively controls the loading conditions. A high value for 
$g$ amounts to a large angle of contact between the materials, or 
slow loading conditions, and vice versa. As long as $1/g$ is in a 
range where it allows the front to develop but not extend out of 
the system, we see no change in any scaling exponents we have 
investigated. This behavior is supported by the experimental 
results where no dependence on the loading conditions was found 
and the average propagation velocity of the front spanned more than 
four and a half orders of magnitude \cite{ttssm11}. 

The parameter $e$ controls the material behavior. A small value for 
$e$ can amount to model a system of low modulus of elasticity or 
a long length scale, and vice versa. Based on the behavior alluded to
in our model, we may propose the following scenario.
For low values of $e$, the fracture process is due to stress concentration 
leading to damage forming on the fracture front. This regime is classified 
by a roughness exponent of $\zeta_{\textrm{large}}=0.39$ and an 
underlying burst process of same exponent $H=\zeta_{\textrm{large}}$. 
As $e$ increases, the burst process changes behavior to 
$H=\zeta_{\textrm{small}}=0.67$, while the front is still 
characterized by $\zeta_{\textrm{large}}$. As $e$ is increased 
further, the effect of the burst exponent begins to show on the 
scale of the front and damage begins to form ahead of the front. 
At this point, the roughness exponent changes from 0.39 and 
grows into 0.67. At higher $e$, both the burst process and the 
roughness of the front is now characterized by the larger exponent 
$H=\zeta_{\textrm{small}}=0.67$.  

\section{Conclusion}

We have developed a minimal model which we have previously shown to 
contain two different fracture growth processes which reproduce 
results from experiments with fracture fronts of roughness 
$\zeta_{\textrm{small}}=0.67$ on small length scales and 
$\zeta_{\textrm{large}}=0.39$ on larger length scales \cite{gsh13}. 
In this paper we have shown that this quasistatic model also 
produces data consistent with the experimental 
analysis of the dynamics of the 
fracture front. Most notably we recover the existence and behavior 
of a depinning and a pinning regime. In the pinning regime we recover 
the same 0.39-scaling exponent of the front in the bursts. In the 
depinning regime we can see traces of this behavior, but more importantly, 
we recover the 0.67-scaling exponent of the front seen on shorter 
length scales. From this, we conjecture that the collective behavior 
of the underlying bond breaking process is first and most 
visible in the damage bursts. The global front roughness changes 
according to the change in burst process when one process 
suppresses the other, i.e., on a different scale.

\section*{Acknowledgments}

We are grateful for discussions and suggestions by D.\ Bonamy, K.\
J.\ M{\aa}l{\o}y, L.\ Ponson and K.\ T.\ Tallakstad.
This work was supported by the Norwegian Research Council (NFR) under 
grant number 177591/V30.
Part of this work made use of the facilities of HECToR, the UK's
national high performance computing service, which is
provided by UoE HPCx Ltd at the University of
Edinburgh, Cray Inc and NAG Ltd, and funded by
the Office of Science and Technology through
EPSRC's High End Computing Programme.
\bibliographystyle{plain}

\begin{thebibliography}{11}

\bibitem{l93} B.\ r.\ Lawn, {\it Fracture of Brittle Solids,\/} 2nd Ed.\
(Cambridge Univ.\ Press, Cambridge, 1993).

\bibitem{srvm95} J.\ Schmittbuhl, S.\ Roux, J.-P.\ Vilotte and K.\ J.\
M{\aa}l{\o}y, Phys.\ Rev.\ Lett.\  {\bf 74} 1787 (1995).

\bibitem{sm97} J.\ Schmittbuhl and K.\ J.\ M{\aa}l{\o}y, Phys.\ Rev.\ Lett.\
{\bf 78} 3888 (1997).

\bibitem{ref97} S.\ Ramanathan, D.\ Ertas and D.\ S.\  Fisher, Phys.\
Rev.\ Lett.\ {\bf 79}, 873 (1997).

\bibitem{dsm99} A.\ Delaplace, J.\ Schmittbuhl and K.\ J.\ M{\aa}l{\o}y,
Phys.\ Rev.\ E.\  {\bf 60} 1337 (1999).

\bibitem{rk02} A.\ Rosso and W.\ Krauth, Phys.\ Rev.\ E.\ {\bf 65}
R025101 (2002).

\bibitem{shb03} J.\ Schmittbuhl, A.\ Hansen and G.\ G.\ Batrouni, Phys.\
Rev.\ Lett.\ {\bf 90} 045505 (2003).

\bibitem{sgtshm10} S.\ Santucci, M.\ Grob, R.\ Toussaint, J.\ Schmittbuhl,
A.\ Hansen and K.\ J.\ M{\aa}l{\o}y, Europhys.\ Lett.\  {\bf92} 44001 (2010).

\bibitem{b09} D.\ Bonamy, J.\ Appl.\ Phys.\ D, {\bf 42}, 214014 (2009).

\bibitem{bb11} D.\ Bonamy and E.\ Bouchaud, Phys.\ Rep.\ {\bf 498} 1 (2011).

\bibitem{ms01} K.\ J.\ M{\aa}l{\o}y and J.\ Schmittbuhl, Phys.\ Rev.\
Lett.\ {\bf 87}, 105502 (2001).

\bibitem{msst06} K.\ J.\ M{\aa}l{\o}y, S.\ Santucci, J.\ Schmittbuhl
and R.\ Toussaint, Phys.\ Rev.\ Lett.\ {\bf 96}, 045501 (2006).

\bibitem{ttssm11} K.\ T.\ Tallakstad, R.\ Toussaint, S.\ Santucci,
J.\ Schmittbuhl and K.\ J.\ M{\aa}l{\o}y, Phys.\ Rev.\ E, {\bf 83},
046108 (2011).

\bibitem{ltseatsm11} O.\ Lenglin{\'e}, R.\ Toussaint, J.\ Schmittbuhl,
J.\ E.\ Elkhoury, J.\ P.\ Ampuero, K.\ T.\ Tallakstad, S.\ Santucci and
K.\ J.\ M{\aa}l{\o}y, Phys.\ Rev.\ E, {\bf 84}, 036104 (2011).

\bibitem{bblp93} J.\ P.\ Bouchaud, E.\ Bouchaud, G.\ Lapasset and
J.\ Plan{\`e}s, Phys.\ Rev.\ Lett.\ {\bf 71} 2240 (1993).

\bibitem{lsz10} L.\ Laurson, S.\ Santucci and S.\ Zapperi, Phys.\ Rev.\ E,
{\bf 81}, 046116 (2010).

\bibitem{bbfrr02} E.\ Bouchaud, J.\ P.\ Bouchaud, D.\ S.\ Fisher, S.\
Ramanathan and J.\ R.\ Rice, J.\ Mech.\ Phys.\ Sol.\ {\bf 50}, 1703 (2002).

\bibitem{bhs02} G.\ G.\ Batrouni, A.\ Hansen and J.\ Schmittbuhl,
Phys.\ Rev.\ E.\ {\bf 65} 036126 (2002).

\bibitem{gsh12} K.\ S.\ Gjerden, A.\ Stormo and  A.\ Hansen,
J.\ Phys: Conf.\ Series, {\bf 402}, 012039 (2012).

\bibitem{gsh13} K.\ S.\ Gjerden, A.\ Stormo and A.\ Hansen, 
arXiv:1301.2174 (2013).

\bibitem{hbrs07} A.\ Hansen, G.\ G.\ Batrouni, T.\ Ramstad and
J.\ Schmittbuhl,  Phys.\ Rev.\ E.\  {\bf 75} 030102 (2007).

\bibitem{bsp08} D.\ Bonamy, S.\ Santucci and L.\ Ponson, Phys.\
Rev.\ Lett.\ {\bf 101}, 045501 (2008).

\bibitem{sdm01} J.\ Sethna, K.\ Dahmen and C.\ Myers, Nature, {\bf 410},
242 (2001).

\bibitem{mrs97} A.\ R.\ Mehrabi, H.\ Rassamdana, and M.\ Sahimi, Phys.\
Rev.\ E {\bf 56} 712 (1997).

\bibitem{shn98} I.\ Simonsen, A.\ Hansen and O.\ M.\ Nes, Phys.\ Rev.\
E {\bf 58} 2779 (1998).

\bibitem{phc10} S.\ Pradhan, A.\ Hansen and B.\ K.\ Chakrabarti,
Rev.\ Mod.\ Phys.\ {\bf 82}, 499 (2010).

\bibitem{j85} K.\ L.\ Johnson, {\it Contact Mechanics\/} (Cambridge
University Press, Cambridge, 1985).

\bibitem{bhn86} G.\ G.\ Batrouni, A.\ Hansen and M.\ Nelkin, Phys.\
Rev.\ Lett.\ {\bf 57}, 1336 (1986).

\bibitem{bh88} G.\ G.\ Batrouni and A.\ Hansen, J.\ Stat.\ Phys.\
{\bf 52}, 747 (1988).

\bibitem{sgh12} A.\ Stormo, K.\ S.\ Gjerden and A.\ Hansen,
Phys.\ Rev.\ E, {\bf 86}, R025101 (2012).

\bibitem{h05} A.\ Hansen, Comp.\ Sci.\ Engn.\ {\bf 7}, (5), 90 (2005).

\bibitem{srg85} B.\ Sapoval, M.\ Rosso and J.\ -F.\ Gouyet, J.\ Phys.\ Lett.\
{\bf 46}, L149 (1985).

\bibitem{gr05} J.-F.\ Gouyet and M.\ Rosso, Physica A {\bf 357} 86 (2005).

\bibitem{drp01} A.\ Delaplace, S.\ Roux and G.\ Pijaudier-Cabot,
J.\ Engn.\ Mech.\ {\bf 127} 646 (2001).

\bibitem{fv85} F.\ Family and T.\ Vicsek, J.\ Phys.\ A {\bf 18}, L75
(1985).

\bibitem{sa92} D.\ Stauffer and A.\ Aharony, {\it Introduction to
percolation theory\/} (Francis and Taylor, London, 1992).

\bibitem{m77} B.\ B.\ Mandelbrot, {\it Fractal forms, chance and
dimensions\/} (W.\ H.\ Freeman, San Francisco, 1977).

\bibitem{sdm01-1} J.\ Schmittbuhl, A.\ Delaplace and K.\ J.\
M{\aa}l{\o}y in E.\ Bouchaud et al.\ (Eds.) {\it Physical Aspects of
Fracture\/} (Kluwer Acad.\ Publ., Amsterdam, 2001).

\bibitem{dk07} O.\ Duemmer and W.\ Krauth, J.\ Stat.\ Phys.\ P01019 (2007).

\bibitem{e43} A.\ de Saint-Exup{\'e}ry, {\it Le Petit Prince\/} 
({\'E}ditions Gallimard, Paris, 1943).

\end{thebibliography}

\end{document}